\documentclass[prb,twocolumn, aps,letterpaper, 
reprint, %endfloats,
preprintnumbers, superscriptaddress
]{revtex4-2}

\usepackage{graphicx}% Include figure files
\usepackage{dcolumn}% Align table columns on decimal point
\usepackage{bm}% bold math
\usepackage{lineno}

\usepackage {multirow}

\usepackage{soul} %skip \ref and \cite in \hl
\soulregister\cite7
\soulregister\ref7
\soulregister\citep7
\soulregister\citet7
\soulregister\pageref7

\usepackage{amsthm,amsmath,amssymb}
\usepackage{mathrsfs}
\usepackage{appendix}
\usepackage{amstext} 
\usepackage{url}
\usepackage{float} 
\usepackage[usenames,dvipsnames]{color}
\usepackage[colorlinks=true,citecolor=blue,urlcolor=black]{hyperref}% add hypertext capabilities
\usepackage{dcolumn}% Align table columns on decimal point
\usepackage{booktabs}
\usepackage{graphicx}
\usepackage[linesnumbered,algoruled,boxed,lined]{algorithm2e} 
\usepackage{siunitx}
\usepackage{makecell}

\newcommand{\be}{\begin{equation}}
\newcommand{\ee}{\end{equation}}

\newcommand{\sket}[1]{{\ensuremath{\lvert#1\rangle}}}
\newcommand{\lket}[1]{{\ensuremath{\left\lvert#1\right\rangle}}}
\newcommand{\ket}[1]{\if@display\lket{#1}\else\sket{#1}\fi}

\usepackage{stackengine}
\newcommand{\sbra}[1]{{\ensuremath{\langle#1\rvert}}}
\newcommand{\lbra}[1]{{\ensuremath{\left\langle#1\right\rvert}}}
\newcommand{\bra}[1]{\if@display\lbra{#1}\else\sbra{#1}\fi}

\newcommand{\sketbra}[2]{{\ensuremath{\lvert #1\rangle\!\langle #2\rvert}d}}
\newcommand{\lketbra}[2]{{\ensuremath{\left\lvert #1\right\rangle\!\!\left\langle #2\right\rvert}}}
\newcommand{\ketbra}[2]{\if@display\lketbra{#1}{#2}\else\sketbra{#1}{#2}\fi}

\newcommand{\zy}{\textcolor{black}}

\theoremstyle{plain}

\theoremstyle{definition}

\begin{document}

\title{Independent Optical Frequency Combs Powered\\ 546~km Field Test of Twin-Field Quantum Key Distribution}

\begin{abstract}
Owing to its repeater-like rate-loss scaling, twin-field quantum key distribution (TF-QKD) has repeatedly exhibited in laboratory its superiority for secure communication
over record fiber lengths.  
Field trials pose a new set of challenges however, which must be addressed before the technology's roll-out into real-world.
Here, we verify in field the viability of using independent optical frequency combs---installed at sites separated by a straight-line distance of 300~km---to achieve  a versatile TF-QKD setup that has no need for optical frequency dissemination and thus enables an open and network-friendly fiber configuration.  
Over 546 and 603 km symmetric links, we record a finite-size secure key rate (SKR) of 0.53~bit/s and an asymptotic SKR of 0.12 bit/s, respectively. 
Of practical importance,  the setup is demonstrated to support 44~km fiber 
asymmetry in the 452 km link.  Our work marks an important step towards incorporation of long-haul fiber links into large quantum networks. 
\end{abstract}

\author{Lai~Zhou}
\thanks{Equal contribution.}
\author{Jinping~Lin}
\author{Chengfang~Ge}
\affiliation{Beijing Academy of Quantum Information Sciences, Beijing 100193, China} 
\author{Yuanbin~Fan}
\affiliation{Beijing Academy of Quantum Information Sciences, Beijing 100193, China} 
\author{Zhiliang~Yuan}
\email{yuanzl@baqis.ac.cn}
\affiliation{Beijing Academy of Quantum Information Sciences, Beijing 100193, China}

\makeatletter
    % store affiliation, author macro definitions
    % before being cleared by first \maketitle
    \let\tmpaffiliation\affiliation
    \let\tmpauthor\author
    \let\tmpthanks\thanks
    % store abstract macro definition before temporary clearing it
    % to prevent abstract being printed after first author block
    \let\tmpabstract\frontmatter@abstract@produce
    \let\frontmatter@abstract@produce\relax
    % prevent vertical space being added after first author block
    \let\frontmatter@finalspace\relax
    % print title and first author block
    \maketitle
    
    % restore definition of vertical space at end of author block
    \def\frontmatter@finalspace{\addvspace{18\p@}}
    % restore definitions of \maketitle, affiliation, author, abstract
    \let\maketitle\frontmatter@maketitle
    \let\affiliation\tmpaffiliation
    \let\author\tmpauthor
     \let\thanks\tmpthanks
    \let\frontmatter@abstract@produce\tmpabstract
    % prevent printing title a second time
    \let\frontmatter@title@produce\relax
    \makeatother

\author{Hao~Dong}
\thanks{Equal contribution.}
\affiliation{Hefei National Research Center for Physical Sciences at the Microscale and School of Physical Sciences, University of Science and Technology of China, Hefei, 230026, Anhui, China} 
\affiliation{Jinan Institute of Quantum Technology and CAS Center for Excellence in Quantum Information and Quantum Physics, University of Science and Technology of China, Jinan, 250101, Shandong, China}
\author{Yang~Liu}
\affiliation{Jinan Institute of Quantum Technology and CAS Center for Excellence in Quantum Information and Quantum Physics, University of Science and Technology of China, Jinan, 250101, Shandong, China}
\affiliation{Hefei National Laboratory, University of Science and Technology of China, Hefei, 230088, Anhui, China}
\author{Di~Ma}
\affiliation{Jinan Institute of Quantum Technology and CAS Center for Excellence in Quantum Information and Quantum Physics, University of Science and Technology of China, Jinan, 250101, Shandong, China}
\author{Jiu-Peng~Chen}
\affiliation{Hefei National Research Center for Physical Sciences at the Microscale and School of Physical Sciences, University of Science and Technology of China, Hefei, 230026, Anhui, China} 
\affiliation{Jinan Institute of Quantum Technology and CAS Center for Excellence in Quantum Information and Quantum Physics, University of Science and Technology of China, Jinan, 250101, Shandong, China}
\affiliation{Hefei National Laboratory, University of Science and Technology of China, Hefei, 230088, Anhui, China} 
\author{Cong~Jiang}
\affiliation{Jinan Institute of Quantum Technology and CAS Center for Excellence in Quantum Information and Quantum Physics, University of Science and Technology of China, Jinan, 250101, Shandong, China}
\affiliation{Hefei National Laboratory, University of Science and Technology of China, Hefei, 230088, Anhui, China}
\author{Xiang-Bin~Wang}
\affiliation{Jinan Institute of Quantum Technology and CAS Center for Excellence in Quantum Information and Quantum Physics, University of Science and Technology of China, Jinan, 250101, Shandong, China}
\affiliation{State Key Laboratory of Low Dimensional Quantum Physics, Department of Physics, Tsinghua University, Beijing 100084, China}
\affiliation{Hefei National Laboratory, University of Science and Technology of China, Hefei, 230088, Anhui, China} 
\author{Li-Xing~You}
\affiliation{Shanghai Key Laboratory of Superconductor Integrated Circuit Technology, Shanghai Institute of Microsystem and Information Technology, Chinese Academy of Sciences, Shanghai 200050, China}
\author{Qiang~Zhang}
\affiliation{Hefei National Research Center for Physical Sciences at the Microscale and School of Physical Sciences, University of Science and Technology of China, Hefei, 230026, Anhui, China} 
\affiliation{Jinan Institute of Quantum Technology and CAS Center for Excellence in Quantum Information and Quantum Physics, University of Science and Technology of China, Jinan, 250101, Shandong, China}
\affiliation{Hefei National Laboratory, University of Science and Technology of China, Hefei, 230088, Anhui, China} 
\author{Jian-Wei~Pan}
\affiliation{Hefei National Research Center for Physical Sciences at the Microscale and School of Physical Sciences, University of Science and Technology of China, Hefei, 230026, Anhui, China} 
\affiliation{Hefei National Laboratory, University of Science and Technology of China, Hefei, 230088, Anhui, China} 
\date{\today}

\maketitle

\section {Introduction}

Unveiled in 1984~\cite{bennett2014quantum}, quantum key distribution (QKD) is a revolutionary means for exchanging cryptographic keys over fiber networks~\cite{peev2009secoqc,sasaki2011field,dynes2019cambridge,chen2021integrated,ribezzo2023,bersin2024}. Forty years later, it has emerged as a practical, and arguably the only known future-proof, candidate to mitigate imminent threats posed by quantum computers towards existing public-key  infrastructures.  
The technology has been demonstrated in laboratory capable of delivering 10-100~Mbit/s secure key rates (SKR's)~\cite{yuan201810,li2023high,Grunenfelder2023} and long-haul reach up to 421~km over a continuous fiber link~\cite{boaron2018secure}. 
Excitingly, using elegant ideas of twin-field (TF)~\cite{Lucamarini2018} or post-measurement pairing~\cite{zeng2022mode,xie2022breaking},
fiber distances exceeding 500~km can now be routinely achieved~\cite{chen2020sending,pittaluga21,wang22,zhou2023quantum,liu2023experimental,zhou2023experimental} while achieving measurement-device-independent (MDI) security~\cite{lo2012measurement}. 

Long-distance links are highly desirable for deployment in a large-scale QKD network because they can reduce the number of trusted nodes and enhance overall network security.  
To this end, TF-QKD is an attractive choice. Its viability for field use has recently been tested in early trials~\cite{liu2021field,chen2021twin,clivati2022}, with the longest fiber of 511~km linking two cities~\cite{chen2021twin}.  These trials successfully overcome TF-QKD's technical challenge for stabilizing and/or tracking the differential phase between laser fields transmitted from remote locations. However, the underlying solutions require either the communication users cloning a common laser frequency~\cite{liu2021field,clivati2022} or calibrating their laser frequencies at regular intervals~\cite{chen2021twin}. An auxiliary channel for frequency dissemination is then necessary, which results to a closed-loop fiber configuration and will thus restrict flexibility towards a scalable and switchable network. 

Optical frequency comb (OFC) is emerging as a powerful technique for TF-QKD~\cite{clivati2022,zhou2023quantum}. 
Inter-wavelength coherence it brings  enables dual-band stabilization~\cite{pittaluga21} to function over fiber asymmetry of 10's kilometers, as field-trialed in a closed-loop fiber configuration~\cite{clivati2022}. 
Further, adoption of electro-optic combs removes the need for optical frequency dissemination and allows TF-QKD's operation over an open quantum channel~\cite{zhou2023quantum}, thus 
reducing its fiber complexity to the same level as MDI-QKD~\cite{lo2012measurement}. However, TF-QKD over an open quantum channel has yet to be demonstrated in a field environment nor has an OFC-powered setup delivered a positive SKR.

Here, we place the OFC powered technology under stringent field test and experimentally verify the viability of the TF-QKD setup we implemented to operate across a deployed 427~km link. 
%With added fiber spools 
Over symmetric quantum channels achieved by adding extra fiber spools, the setup is verified capable of delivering 
a finite-size SKR of 0.53~bit/s at 546~km and an asymptotic SKR of 0.12~bit/s at 603~km using the efficient TF-QKD variant of sending-not-sending (SNS) protocol~\cite{wang2018twin,hu2019sending}. 
Furthermore, we demonstrate that the open channel scheme supports a large fiber asymmetry of 44~km in the 452~km asymmetric setup. This work proves TF-QKD's operation without optical frequency dissemination in field and thus paves the way towards incorporation of long-haul links into secure quantum network.

\begin{figure*}[t]
\centering
\includegraphics[width=1.8\columnwidth]
{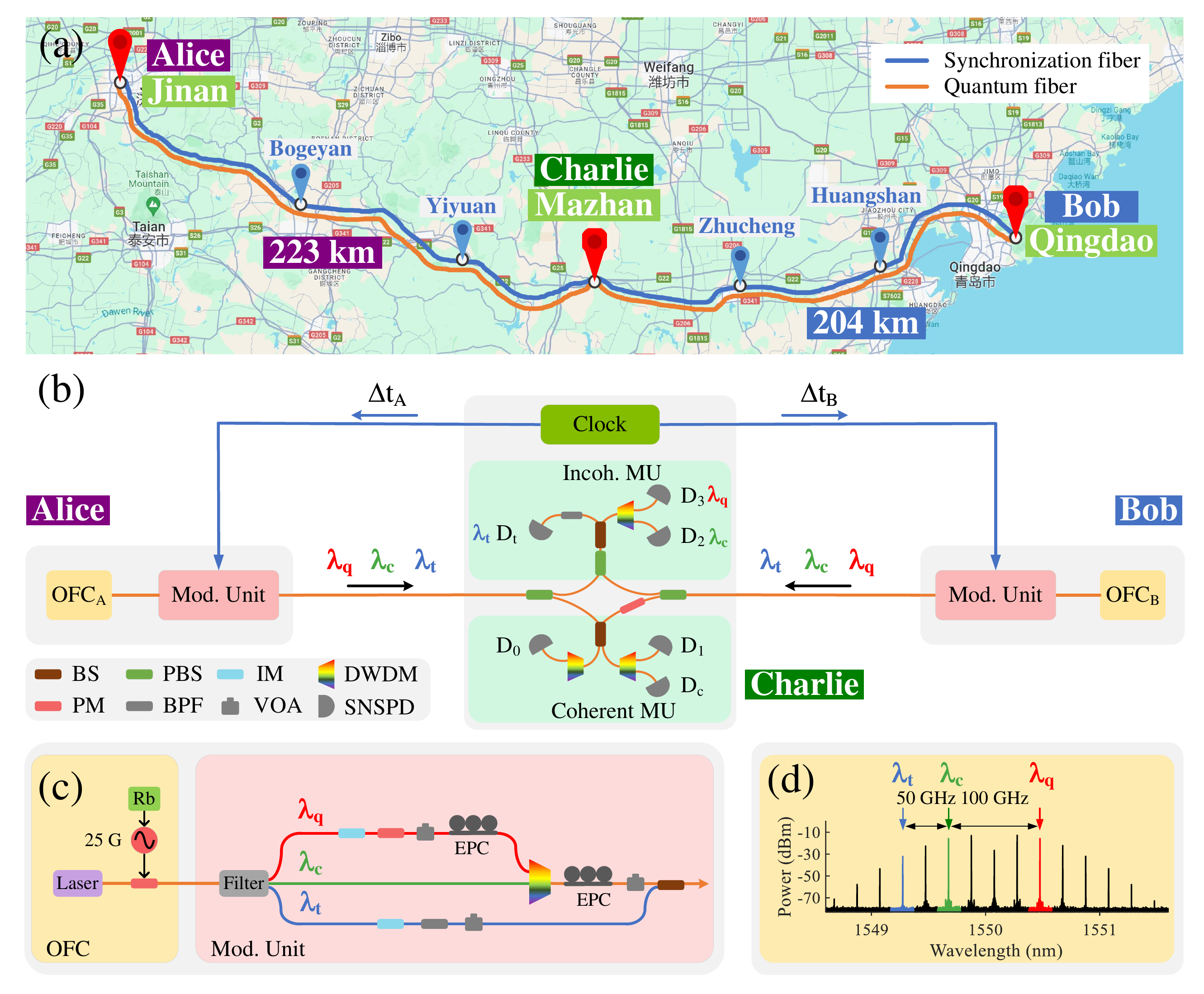}	\caption{Field test setup. (a) Deployed fiber route.  (b) TF-QKD setup. 
(c) Transmitter, including OFC and Modulation Unit.  
(d) OFC spectrum. IM, intensity modulator; PM, phase modulator; BS, beam splitter; PBS, polarization beam splitter; DWDM, dense wavelength division multiplexer; SNSPD: superconducting nanowire single photon detector; EPC, electrically driven polarization controller; VOA, variable optical attenuator; BPF, band pass filter.}
\label{fig_experiment_setup}
\end{figure*}

\section{Field experimental setup}

As shown in Fig.~\ref{fig_experiment_setup}(a), the experimental setup is installed across three geographical sites connected by commercial fiber.  The communication users (referred to as Alice and Bob) are placed at opposite ends of the fiber link, i.e., the metropolitan cities of Jinan and Qingdao, while the measurement node Charlie situates at Mazhan, a town that lies in the middle. 
\zy{The straight-line distance between Alice and Bob is 300~km.}
Two parallel fibers are allocated. 
The quantum fiber carrying the quantum signal has a total length of 427~km and is characterised to have a loss coefficient of 0.190 (0.185) ~dB~km$^{-1}$ between Charlie and Alice (Bob).
The synchronization fiber of the same length is dedicated for transmission of Charlie's clock signal to time the users' modulation units, see  Fig.~\ref{fig_experiment_setup}(b).  Optical amplifiers are used to boost the clock signal in the synchronization fiber. 
Internet handles all other classical communications, which are not jitter-critical. 
To test quantum channels longer than 427~km, we add fiber spools (G654.C fiber) with a typical loss coefficient of 0.160~dB~km$^{-1}$ at Charlie's site. 

Each user owns an independent OFC, made through 25~GHz microwave modulation to an ultra stable laser's output, see Figs.~\ref{fig_experiment_setup}(b, c).  
Precise comb spacing,  
crucial for this TF-QKD implementation, is ensured by referencing each 25~GHz microwave driver to a local Rubidium clock of an accuracy of $5 \times 10^{-11}$. 
Each OFC's laser is locked to a high-fineness cavity to have a sub-Hertz short-term linewidth using Pound-Drever-Hall technique.  Due to referencing to stable, independent cavities,  two lasers are characterized to drift at a dominantly linear rate of 1.8~kHz/h, which is compensated for by adjusting the frequency offset in the locking path of Alice's laser.  

Each OFC generates 15 spectral lines, see Fig.~\ref{fig_experiment_setup}(d).
We choose 3 lines,  in alignment with the ITU 50~GHz DWDM  
Grid,  for carrying quantum ($\lambda_q=1550.495$~nm, Ch 33.5), channel stabilization ($\lambda_c=1549.694$~nm, Ch 34.5) and timing ($\lambda_t=1549.293$~nm, Ch 35) signals, respectively.
The $\lambda_q$ light 
is carved and modulated into a sequence of blocks of 200~ns each, \textcolor{black}{see Fig.~\ref{fig:encoding_block} in Appendix~B}.  Within each block, the first 100~ns comprises a train of 300~ps pulses clocked at 1~ns intervals and encoded  to the single photon level to form the `quantum signal', 
while in the remaining 100~ns centers one broad pulse of 70~ns duration serving as the `weak quantum reference'. 
The effective clock rate for quantum signal transmission is 500~MHz. 
The $\lambda_c$ light is continuous-wave and serves as the `strong channel reference', while the $\lambda_t$ signal is carved into 10~MHz, 15~ns pulses and used as `time reference' for Charlie to determine photon arrivals by his incoherent measurement unit (Fig.~\ref{fig_experiment_setup}(b)). 

Alice and Bob's modulation units are timed with respective dynamical delays ($\Delta t_A$, $\Delta t_B$) to produce their quantum ($\lambda_q$) and time reference ($\lambda_t$) pulses by the incoming 50~MHz clock signal transmitted by Charlie through the synchronization fiber, \textcolor{black}{see Fig.~\ref{fig:clock} in Appendix~B.} Its wavelength is chosen to be $1559.79$~nm, spectrally distant from $\lambda_q$, to avoid cross-fiber noise
~\cite{chen2021twin,liu2021field}. 
Polarizations of each user's $\lambda_c$ and $\lambda_q$ signals are independently
pre-compensated for by minimising the count rate at detectors $D_2$ and $D_3$ by adjusting the electric polarization controllers (EPCs) at the transmitter units,
so that they are predominantly transmitted to Charlie's coherent measurement unit through a polarization beam splitter (PBS), see Figs.~\ref{fig_experiment_setup}(b,c).
In the coherent measurement unit, the interference outcomes are demultiplexed in wavelength before entering detectors $D_0$, $D_1$ and $D_c$. For more information on the experimental setup, refer to \textcolor{black}{Appendix~B}. 

\section{Field trial preparation}

The first task for the field trial is to align the OFC lasers
to have their frequency difference within a confirmed active phase compensation bandwidth of $\pm 5$~kHz. 
Using a fast photodiode and a frequency spectrum analyzer, we measure the beat frequency between the lasers at Mazhan after transmission from Jinan and Qingdao.  Here, the users' comb generation and modulation units are temporarily bypassed so as to ensure sufficient optical power entering the fast photodiode.  Limited by fiber length fluctuation, we measure the beating frequency within an accuracy of $\sim 2$~kHz. This coarse result is sufficient for subsequent real-time frequency tracking via single photon detection. 

\begin{figure*}[t]
\newpage
\centering
\includegraphics[width=1.6\columnwidth]{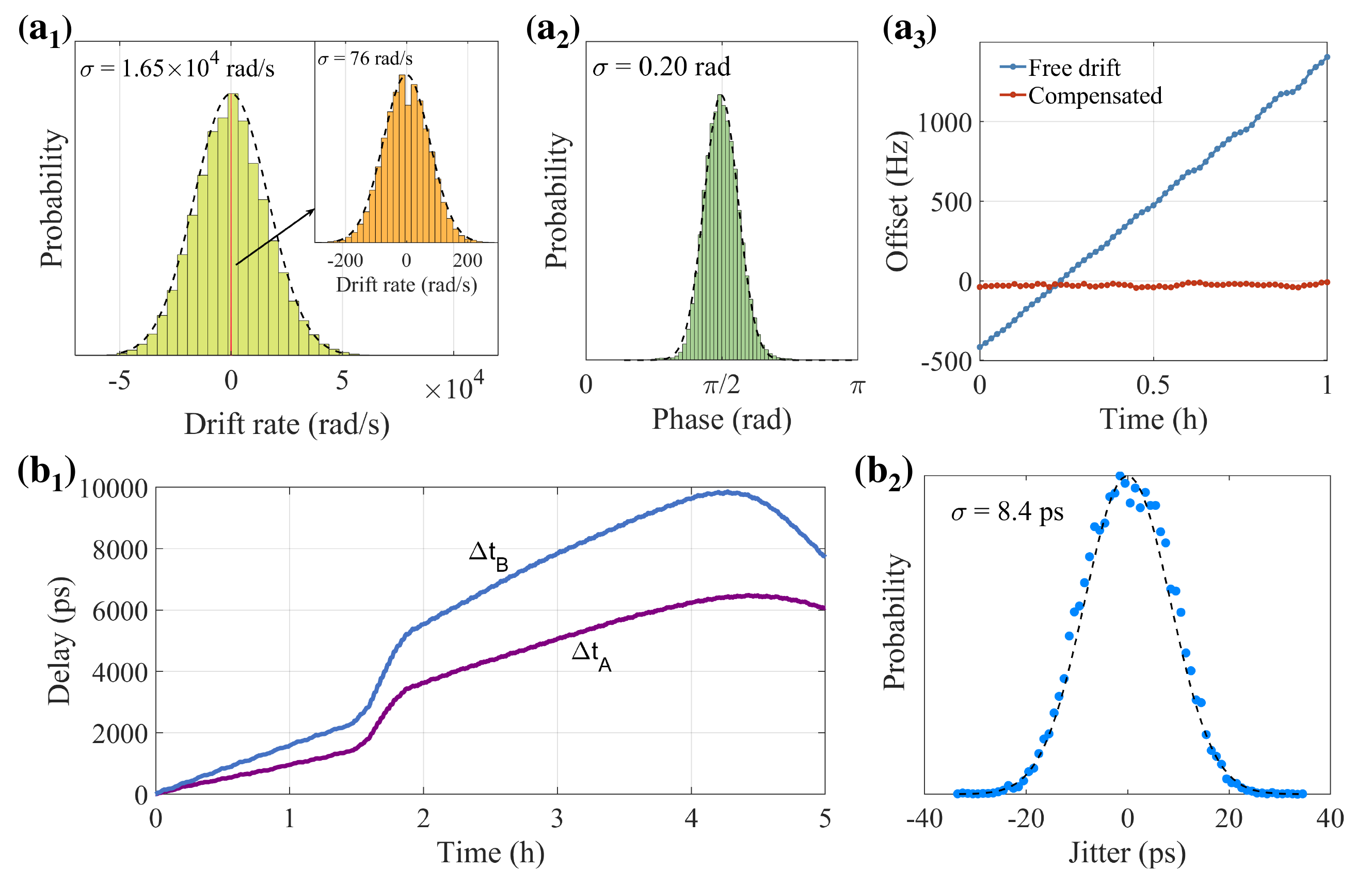}
	\caption{Autonomous alignment for phase, optical frequency and arrival time. 
 \textbf{a}, Coherent dual-band phase stabilization results; (a$_1$) Comparison for the phase drift rates of the quantum signal before (light green) and after (orange) turning on the fast phase compensation. 
 Inset: Magnified view of the slowed phase drifts. 
(a$_2$) Phase angle distribution of the $\lambda_q$ light with coherent dual-band stabilization fully on. (a$_3$) Readout of the laser frequency offset without (blue) and with (red) frequency compensation activated.
\textbf{b}, Timing alignment using the incoherent measurement unit (MU). 
(b$_1$) Clock delays to Alice (purple) and Bob (blue); (b$_2$) Jitter distribution of the $\lambda_t$ signals measured by Charlie. All data here were taken over the 427~km deployed fiber. 
}
\label{fig:feedback}
\end{figure*}

We then reinstate the bypassed units into the setup.  
The intensities of Alice and Bob's channel reference $\lambda_c$ signals are set so that they are balanced and jointly produce 6~MHz count rate at Charlie's phase detector ($D_c$ in Fig.~\ref{fig_experiment_setup}(b)).  
To stabilize the differential phase of the $\lambda_c$ signals, a 
proportional–integral–derivative (PID) control module~\cite{zhou2023quantum} is used to adjust the bias to Charlie's phase modulator (PM) so as to stabilize the instantaneous count rate of $D_c$ at 10~$\mu$s intervals.

Figures~\ref{fig:feedback}(a$_1$)  and \ref{fig:feedback}(a$_2$) illustrate the effect of the $\lambda_c$ locking on the phase drift of the quantum \textcolor{black}{signal}.
Without this locking, fiber length fluctuation causes rapid, random drifts in the differential phase of the $\lambda_q$ signal with a standard deviation of $1.65 \times 10^4$~rad/s\textcolor{black}{, see Fig.~\ref{fig:feedback}(a$_1$)}. This rate is similar to previously reported lab results over comparable fiber lengths~\cite{pittaluga21,wang22}. 
Locking the $\lambda_c$ signal at 10~$\mu$s intervals, the $\lambda_q$ phase drift rate is slowed down by a factor of 220 with a standard deviation of $76$~rad/s, see Inset of Fig.~\ref{fig:feedback}(a$_1$). 
The factor is 8.6 times less
than the theoretical prediction~\cite{pittaluga21} of $\lambda_c/|\lambda_c - \lambda_q| \simeq 1900$, and about $7$~times poorer than recent lab operation~\cite{zhou2023quantum}. 
The deterioration arises mainly from the use of \textcolor{black}{two} independent Rubidium clock 
%\st{frequency}
references for OFCs, which have a specified accuracy of $5\times10^{-11}$ each.  With 100~GHz spacing between $\lambda_c$ and $\lambda_q$, this accuracy translates to 44~rad/s \zy{residual phase drift}.
Turning on the $\lambda_q$ phase locking, we measure a Gaussian distribution of 0.20~rad standard deviation for the residual phase, see Fig.~\ref{fig:feedback}(a$_2$). 
This residual phase spreads 1.5~times as much as in the laboratory result~\cite{zhou2023quantum} and contributes 1~$\%$ to the QBER in the X-basis. \textcolor{black}{Please refer to Appendix~B.4.b for details of the active phase stabilization.}

Coherent dual-band stabilization allows real-time readout of the frequency offset based on the compensation signal on the PM~\cite{zhou2023quantum}. 
Fig.~\ref{fig:feedback}(a$_3$) shows two readout results under different experimental conditions: with or without compensation of the linear offset drift turned on. 

With phase and frequency drifts compensated for, the final preparation for the field trial is to maintain temporal alignment between Alice and Bob's quantum pulses.  
Their delays may vary more than 20~ns daily due to environmental temperature surrounding the fiber route. 
To correct, Charlie measures arrival times of Alice and Bob's $\lambda_t$ pulses using detector $D_t$ in his incoherent measurement unit (Fig.~\ref{fig_experiment_setup}(b))  at 1~s intervals and then adjusts the time delay values of $\Delta t_A$ and $\Delta t_B$ accordingly.  
Figure~\ref{fig:feedback}(b$_1$) shows the delays that Charlie dynamically adjusts over a duration of 5~h.  
This dynamical adjustment ensures good temporal overlap between Alice and Bob's quantum signals as inferred from the jitters measured from their $\lambda_t$  \textcolor{black}{pulses}, see  Fig.~\ref{fig:feedback}(b$_2$).
The jitter is measured to be just $8.4$~ps, which is substantially smaller than the 300~ps duration of the quantum pulses and therefore does not exceedingly deteriorate the visibility of the first-order interference.  We measure a visibility of $(97.95 \pm 0.53)~\%$ for the quantum signal 
after coherent dual-band stabilization and transmission over the deployed fiber.

\section {Results}

We implement the sending-not-sending (SNS) TF-QKD protocol using 4 pulse intensities between two encoding bases~\cite{wang2018twin,hu2019sending}.
\textcolor{black}{Each quantum pulse is encoded with a phase value, randomly selected from 16
 phase slices, $\theta \in \{0,\pi/8, 2\pi/8 \cdots 15\pi/8\}$, to meet the requirement of phase randomization and qubit encoding
 in TF-QKD protocol.}
In the Z-Basis, Alice and Bob independently choose at random whether or not send a pulse of a mean photon number of $\mu_Z$ for each transmission time slot, while \textcolor{black}{the other} three pulse intensities ($\mu_0$, $\mu_1$, $\mu_2$, and $\mu_0 < \mu_1 < \mu_2$) are used by each user in the X-Basis in order to apply decoy-state analysis to bound adversaries' information.  
The quantum bit error rate (QBER) in the raw bits, which are formed from events that Alice and Bob both chose the Z-Basis, is unaffected by the channel coherence. Instead, it is proportional to, and thus places a constraint on, the sending probability of $\mu_Z$ pulses.  To address this issue, we apply the actively odd-parity pairing (AOPP) technique~\cite{xu2020sending} in the data post-processing and subsequently use the zig-zag approach~\cite{jiang2020zigzag} to ensure a high SKR with finite-size effects taken into account of. More information on the SNS-AOPP protocol is summarized in \textcolor{black}{Appendix~A}.

We run the SNS-AOPP protocol over three quantum channel lengths. For 500~km or longer, we choose symmetric configuration for optimal performance distance-wise, in which Alice and Bob's transmission distances (losses) to Charlie are strictly matched.
At $546.61$ and $603.87$~km, the link losses are respectively $100.13$ and $108.59$~dB.  
For the shortest distance of $452.46$~km, we opt for a channel asymmetry of 44~km in order to demonstrate TF-QKD's adaptability to real network scenarios. 
At each distance, we transmit at least $4 \times 10^{12}$ quantum pulses, corresponding to 2.22~h experimental time, for obtaining a statistically reliable measurement data set.  For $546.61$~km, the amount of quantum pulses is enlarged 6-fold to $2.772 \times 10^{13}$ in order to rigorously account for finite-size effects.  Protocol settings and measurement results are summarized in \textcolor{black}{Tables~\ref{tab:parameters} and \ref{tab:key rate} in Appendix~C}. 

Figure~\ref{fig:SKR} plots experimentally measured SKR's (red symbols) as a function of quantum channel length, alongside which we include theoretical simulations that were obtained with two parameter sets corresponding to the experiments.  In obtaining the finite-size (solid black) and  asymptotic (dashed) simulation lines, we adopted experimentally measured values of an attenuation coefficient of 0.183~dB/km for the quantum channel and a detection efficiency of 0.66 averaged over \textcolor{black}{detectors $D_0$ and $D_1$.}  
As shown in Fig.~\ref{fig:SKR}, experimental data for the symmetrical experiments are in excellent agreement with their respective simulations. For 603.87~km, the experiment is restricted to the asymptotic case by the access time to the fiber. 

After tests over symmetrical channels, we move to an asymmetric configuration to further verify the robustness of our TF-QKD implementation. In this last trial, just a single 25~km fiber spool is placed into Alice's fiber path on top of the deployed 427~km link, and the distance between Charlie and Alice (Bob) is 248.24 km (204.22 km), extending the link asymmetry to 44~km (9.08~dB) in length (loss).
The length of the spooled fiber is less than 1/4 of those used in the last two experiments, and thus the quantum channel becomes less susceptible to mechanical vibration in Charlie's server room. With help of further securing loose fiber cables, 
the final X-Basis QBER is reduced to 6.87~\% from $>8.0$~\% found in the above symmetrical experiments.  
We obtain an asymptotic SKR of 24.28~bit/s at the distance of 452.46~km, as shown in Fig.~\ref{fig:SKR}. Compared with the theoretical simulation (dashed line), the rate does not exhibit a penalty expected for the fiber asymmetry. This is because the improvement in the X-basis QBER balances out the penalty.
With a transmission of $4.28 \times 10^{12}$ quantum pulses, we obtain a finite-size SKR of 16.06 bit/s. 

To appreciate the advantage of TF-QKD, we compute the absolute repeaterless bound $SKC_0$~\cite{pirandola2017fundamental}, which is the maximum SKR that an ideal  point-point QKD system \zy{clocked at 500~MHz} can offer.
As shown in Fig.~\ref{fig:SKR}, all our experimental data points surpass their respective $SKC_0$ bound values, proving a repeater-like behavior of our system in field environment. At $452.46$ and $546.61$~km, the finite-size SKRs exceed their $SKC_0$ bounds by 6.45 and 7.57 times, respectively.

We compare in Fig.~\ref{fig:SKR} our results with existing state-of-art long-haul field trials.  Decoy-state BB84 feasibility study~\cite{amies2023quantum} or entanglement QKD~\cite{neumann2022continuous} is limited to a distance of 250~km, primarily by the linear rate scaling~\cite{pirandola2017fundamental} of conventional QKD protocols.  On the other hand, TF-QKD field experiments~\cite{liu2021field, chen2021twin} fared substantially longer fiber spans, both exceeding 400~km, thanks to the protocol's intrinsic square-root scaling.
Our setup further extends the communication distances to 546~km (100.13~dB) with finite-size effects included and 603~km (108.59~dB) for a positive asymptotic SKR.  
Moreover, it allows a link asymmetry of 44~km and features an open quantum link that brings ease to deployment.  
Finally, our result constitutes the first field trial that ever breaks 100~dB link loss barrier.

\begin{figure}[t]
\newpage
\centering
\includegraphics[width=\columnwidth]{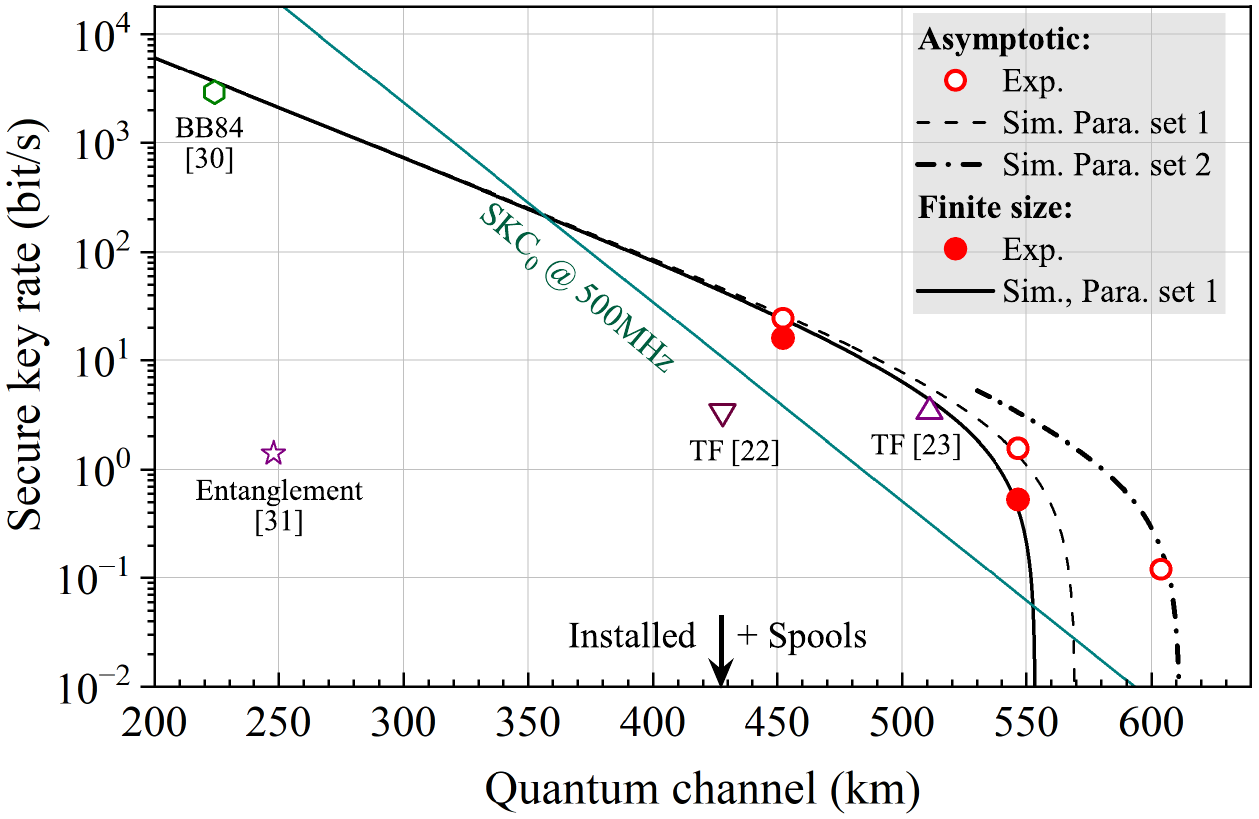}
	\caption{Secure key rate results and simulations. 
According to experiments, two sets of simulation parameters are used. Set 1 (2): 0.660 (0.580) detector efficiency and 0.183 (0.180)~dB/km fiber attenuation. The absolute repeaterless bound $\rm{SKC}_0$ is calculated using 0.183~dB/km fiber attenuation. Previous field trials of decoy-state BB84 feasibility study~\cite{amies2023quantum}, entanglement QKD~\cite{neumann2022continuous} and TF-QKD~\cite{liu2021field,chen2021twin} are shown for comparison. }\label{fig:SKR}
\end{figure}

\section {Conclusion}

We have realized repeater-like quantum communication over
field-deployed fibers of record length and loss.   
Our experiments demonstrate TF-QKD's viability to operate over an open quantum channel in a practical environment as well as its tolerance of channel asymmetry. 
This advance reduces TF-QKD's complexity for real-world deployment and thus makes it an attractive option to upgrade existing QKD networks. 
We believe our techniques will find applications in phase-sensitive quantum repeater and multinode quantum network~\cite{knaut2023entanglement,liu2023multinode}. 

Several advances were made over our recent lab experiments~\cite{zhou2023experimental}, including incorporation of automated time alignment~\cite{chen2021twin} and local clock references for OFCs. Autonomous time, frequency and phase alignment allows the system's sustaining operation at a\textcolor{black}{n effective} clock rate of 500~MHz for quantum signal transmission, while
local references have been demonstrated for the first time in quantum communication.  
In future, the performance of the \textcolor{black}{coherent} dual-band phase stabilization \textcolor{black}{technique} can be upgraded through use of clocks of higher accuracy~\cite{better_clock} 
and/or detectors of higher count rates~\cite{yan2023}.  

\begin{acknowledgments}
This work was supported partially by the National Natural Science Foundation of China (62250710162, 62105034, 12374470, 12174215, T2125010, 61971409, 61971408), the Natural Science Foundation of Beijing (Z230005), the Innovation Program for Quantum Science and Technology (2021ZD0300700, 2023ZD0300100), the National Key R \& D Plan of China (2020YFA0309800), the Key R$\&$D Plan of Shandong Province (2021ZDPT01), and the Shandong Provincial Natural Science Foundation (ZR2022LLZ011). 
H.D., J.P.C., Q.Z. and J.W.P. acknowledge support from the Chinese Academy of Sciences.
Q.Z. acknowledges support from the Taishan Scholar Program of Shandong Province. Z.Y. acknowledges support from the ChangJiang Chair Scholars Program, Ministry of Education.
\end{acknowledgments}

\appendix

\section{Protocol}

In this experiment, we adopt the SNS-TF-QKD protocol~\cite{wang2018twin} with 4-intensities~\cite{yu2019sending}, i.e., there are four phase-randomized weak coherent state (WCS) sources with different intensities in Alice and Bob's side respectively. For clarify, we denote the four sources in Alice's side by $o_a,x_a,y_a,z_a$ and the four sources in Bob's side by $o_b,x_b,y_b,z_b$, whose intensities are $\mu_{a_l}$ ($\mu_{b_l}$) for $l=0,1,2,Z$. Specifically, the sources $o_a,o_b$ are vacuum sources and $\mu_{a_0}=\mu_{b_0}=0$.  To get the highest key rate in the asymmetric channel, the asymmetric 4-intensity SNS-TF-QKD protocol is applied here~\cite{hu2019sending}. Also, in the data post-processing \textcolor{black}{stage}, the  actively odd-parity paring (AOPP) is applied to reduce the bit-flip error rate and further improve the key rate~\cite{jiang2020zigzag,jiang2021composable}.

In this protocol, Alice and Bob would repeat the following process for $N$ times:
\noindent At each time window, Alice (Bob) randomly decides whether it is a decoy window with probability $1-p_{A}$ ($1-p_{B}$), or a signal window with probability $p_{A}$ ($p_{B}$). If it is a signal window, with probability $\epsilon_A$ ($\epsilon_B$), Alice (Bob) chooses the source $z_a$ ($z_b$), and denote it as bit $1$ ($0$); with probability $1-\epsilon_A$ ($1-\epsilon_B$), Alice (Bob) chooses the source $o_a$ ($o_b$), and denote it as bit $0$ ($1$). If it is a decoy window, Alice (Bob) randomly choose the sources $o_a,x_a,y_a$ ($o_b,x_b,y_b$) with probabilities $p_{a_0},p_{a_1},p_{a_2}=1-p_{a_0}-p_{a_1}$ ($p_{b_0},p_{b_1},p_{b_2}=1-p_{b_0}-p_{b_1}$) respectively. To ensure the security of the asymmetric protocol, the following condition is required
\begin{equation}\label{equ:condition}
   \frac{\mu_{a_1}}{ \mu_{b_1}}= \frac{\epsilon_\text{A}(1-\epsilon_\text{B}) \mu_{a_Z}  e^{-\mu_{a_Z}}}{\epsilon_\text{B}(1-\epsilon_\text{A})  \mu_{b_Z} e^{-\mu_{b_Z}}}.
\end{equation}
Then Alice and Bob send their prepared pulses to Charlie who is assumed to perform interferometric measurements on the received pulses and then announces the results to Alice and Bob. If one and only one detector clicks in the measurement process, Charlie also tells Alice and Bob which detector it was, and Alice and Bob take it as a one-detector heralded event. After Alice and Bob repeat the above process for $N$ times, they acquire a series of data, which are used to perform the data post-processing including the AOPP, the decoy-method analysis, the error correction and the privacy amplification to extract the final keys. And they can calculate the secure final key rate by the following formula:
\begin{widetext}
\begin{equation}\label{r2}
\begin{split}
R=&\frac{1}{N}\{n_1^\prime[1-h(e_{1}^{\prime ph})]-fn_t^\prime h(E^\prime)-2\log_2{\frac{2}{\varepsilon_{cor}}}-2\log_2{\frac{1}{\sqrt{2}\varepsilon_{PA}\hat{\varepsilon}}}\},
\end{split}
\end{equation}
\end{widetext}
where $n_1^\prime$ is the number of the untagged bits after AOPP, $e_{1}^{\prime ph}$ is the bit flip error rate of untagged bits after AOPP, $h(x)=-x\log_2x-(1-x)\log_2(1-x)$ is the Shannon entropy, $n_t^\prime$ is the number of the remaining bits after AOPP, $E^\prime$ is the bit-flip error rate of the remaining bits after AOPP, $\varepsilon_{cor}$ is the failure probability of error correction, $\varepsilon_{PA}$ is the failure probability of privacy amplification, and $\hat{\varepsilon}$ is the coefficient while using the chain rules of smooth minimal and maximal entropies~\cite{vitanov2013chain}. The decoy analysis method to get $n_1^\prime$ and $e_{1}^{\prime ph}$ are shown in Ref.~\cite{jiang2021composable}. For simplicity, we do not list the calculation details here.

\section{Detailed description of experimental setup}

\begin{figure*}[t]
\includegraphics[width=1.9\columnwidth]{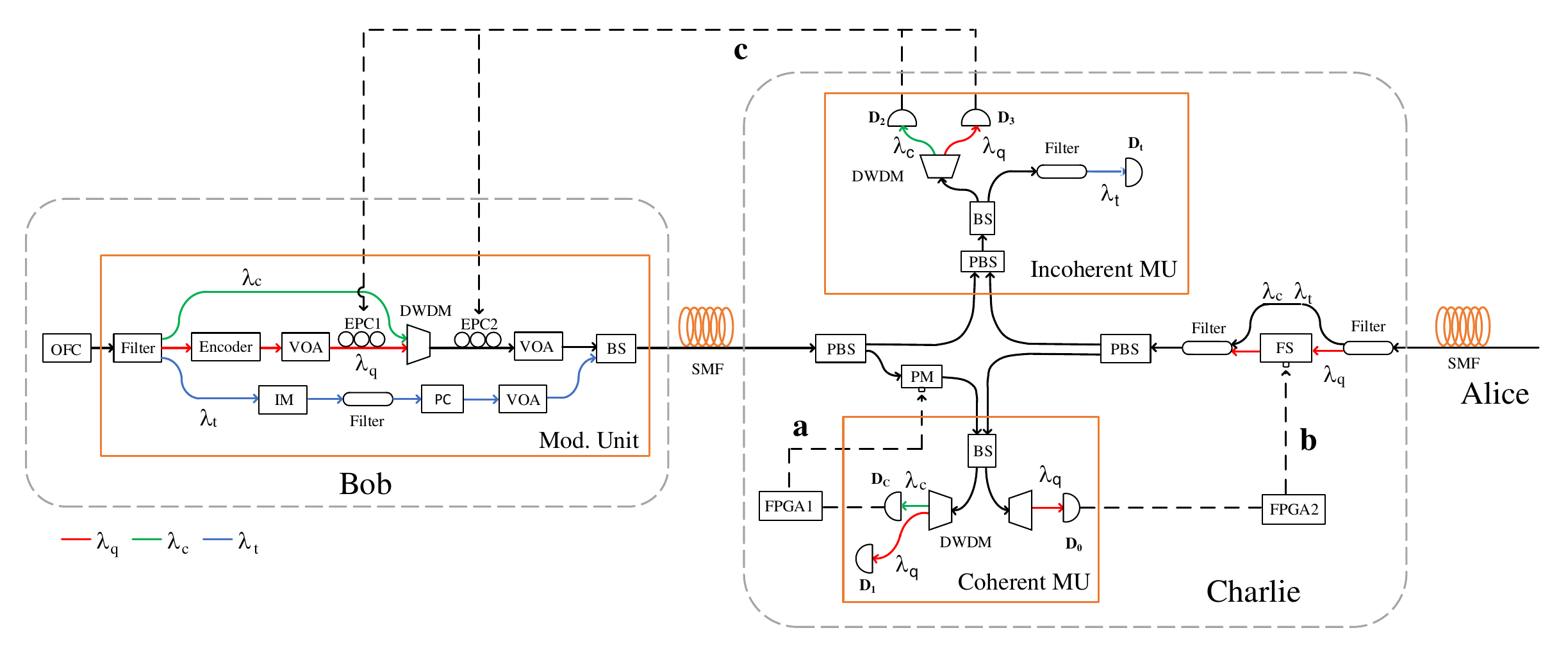}
\caption{Phase and polarization stabilization  setup. \textbf{a}, Fast phase compensation module: FPGA1 is used to detect the counts from detector $D_c$ and derive the error signal driving Charlie's phase modulator (PM) for phase compensation in channel reference $\lambda _c$. 
\textbf{b}, Slow phase compensation module: FPGA2 obtains the error signal from Detector $D_0$ and 
drives the fiber stretch (FS) for compensating the residual phase drift in the quantum reference $\lambda_q$. 
\textbf{c}, Polarization feedback module: EPC1 (EPC2) gets error signal from Detector $D_3$ ($D_2$) and calibrate the polarization drift in wavelength $\lambda_q$ ($\lambda_c$). 
OFC: optical frequency comb; VOA: variable optical attenuator; IM: intensity modulator; BS: beam splitter; SMF: single mode fiber; PBS: polarization beam splitter/combiner; FPGA: field-programmable-gated-array; DWDM: dense wavelength division multiplexing; EPC: electrically driven polarization controller; PC: polarization controller; MU: measurement unit.
}
\label{active feedback}
\end{figure*}

With reference to the schematic shown in Fig.~\ref{active feedback}, we describe in detail our experimental setup with focus on optical frequency comb, modulation units, measurement units, drifts compensation and quantum channel characteristics. Alice and Bob's apparatus are largely identical, so only Bob's setup is shown in Fig.~\ref{active feedback} to avoid repetition.

\subsection{Optical frequency combs}

In the experiment, we use two independent optical frequency combs (OFCs) that are realized through 25~GHz electro-optic modulation to ultra-stable lasers.  
For each OFC, its 25~GHz driver is referenced to a local
Rubidium frequency standard of with a short-term (1~s) stability of $2\times10^{-11}$ and 
an accuracy of $5\times10^{-11}$.  
The frequency accuracy of $5\times 10^{-11}$ causes an error of 5~Hz for 100~GHz span, which means a residual phase drift rate of $7.1$~Hz for the quantum signal ($\lambda_q$) when the classical signal ($\lambda_c$) is perfectly phase stabilized.

\begin{figure}[ht]
\includegraphics[width=.95\columnwidth]{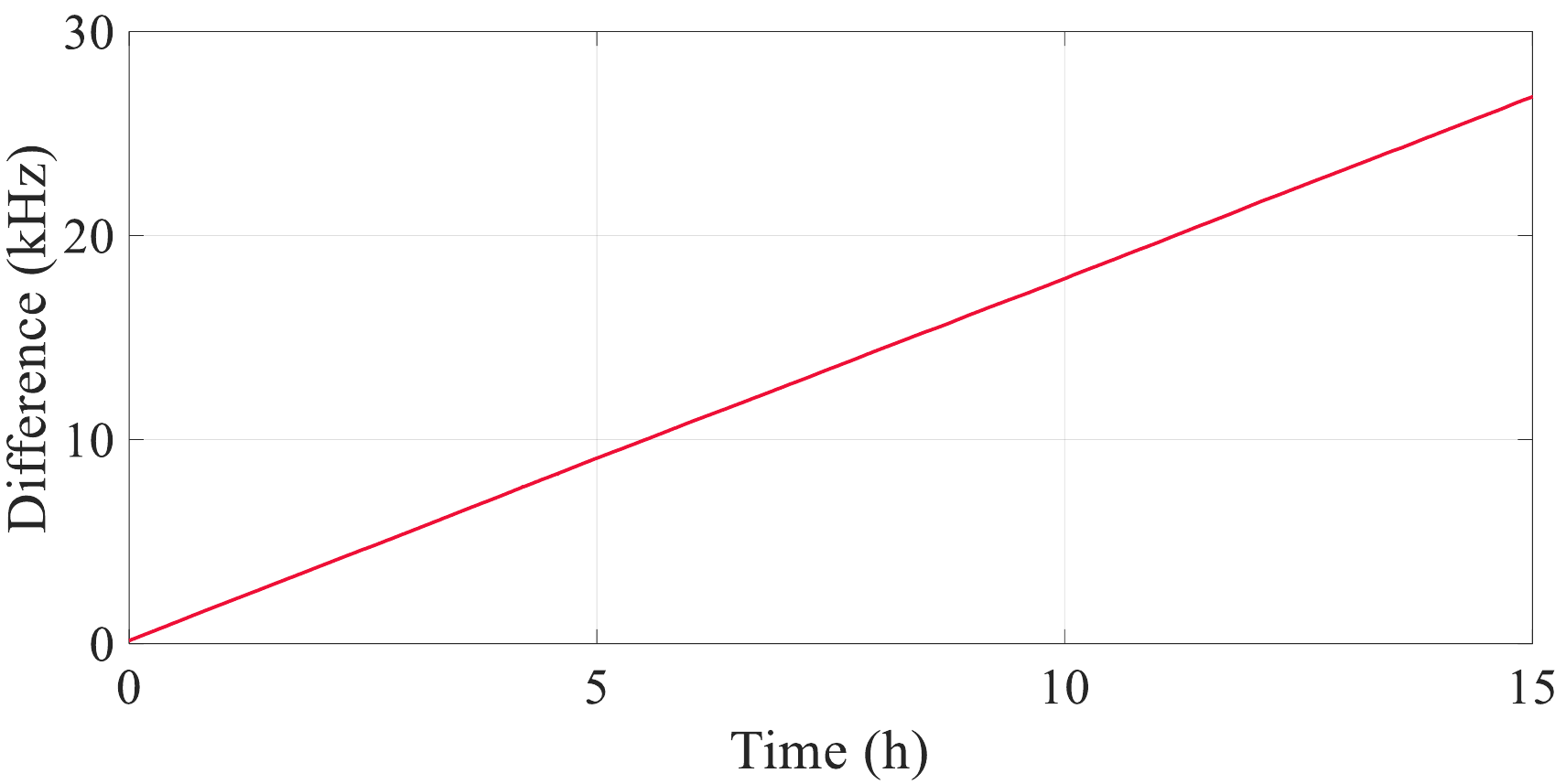}
\caption{Frequency difference between Alice and Bob's ultra-stable lasers. The data was recorded with two lasers placed side by side in laboratory.}
\label{frequency drift}
\end{figure}

The lasers in OFCs were from different manufacturers (Menlo Systems \#ORS-Cubic and Stable Laser Systems \#SLS-INT-1550-200-1), and both feature a sub-Hz short-term linewidth thanks to their use of Pound-Drever-Hall (PDH) technique for locking to a reference cavity. 
Before installation to their respective sites, the two lasers were characterised side by side in laboratory to have a relative linear differential frequency drift of 1777 Hz per hour, see Fig.~\ref{frequency drift}.
The Menlo laser is installed in a research laboratory with network fiber access in Jinan, while the other laser is installed into an equipment cabinet in one of China Unicom's data centers in Qingdao.
The Menlo laser has an electro-optical modulator in its PDH locking path so as to allow frequency tuning over the cavity free spectral range (FSR) and thus alignment of the two lasers' frequencies.
Both lasers are aligned to 1550.094~nm and output 20~mW of power each.

\subsection{Modulation unit}

\subsubsection{Wavelengths arrangement and filtering}

From their electro-optical frequency combs,  each user filters out  
three wavelength lines:  $\lambda_q$ ($1550.495$~nm), $\lambda_c$ ($1549.694$~nm) and $\lambda_t$ ($1549.293$~nm). The quantum wavelength $\lambda_q$ has 100 GHz and 150 GHz spacings from $\lambda_c$ and $\lambda_t$, respectively. $\lambda_q$ passes through the encoder and reunites with $\lambda_c$ using a 50 GHz dense wavelength division multiplexing filter (DWDM). $\lambda_t$ is carved by an intensity modulator (IM) and then goes through a band-pass filter with a bandwidth of 25~GHz for removal of residual $\lambda_q$ light that was leaked through in the first filter. 
The $\lambda_q$ signal \textcolor{black}{is then combined with} the other wavelengths through a beam splitter (BS) before transmission to Charlie through the deployed quantum channel. 
Their intensities are set by three variable optical attenuators (VOAs), and the $\lambda_c$ and $\lambda_t$ signals are set such that they produce 6~MHz and 200~kHz count rates at Charlie, respectively, \textcolor{black}{such that they do not cause noticeable noise contamination to the quantum channel.}

\subsubsection{Encoder}

\begin{figure}[ht]
\includegraphics[width=.9\columnwidth]{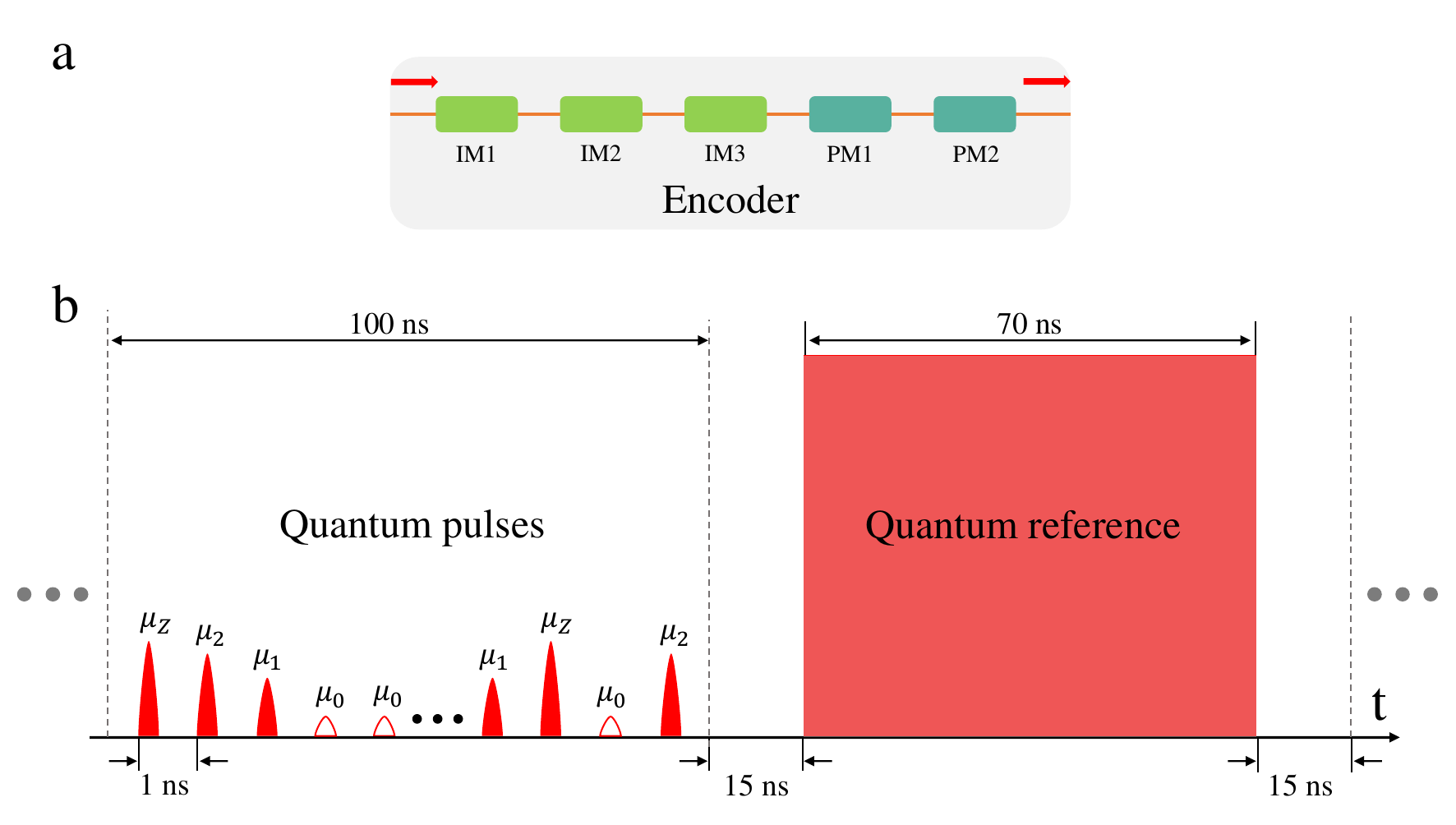}
\caption{Encoder (a) and the encoding sequence of the basic encoding block (b).  
}
\label{fig:encoding_block}
\end{figure}

Alice (Bob) applies block-wise encoding to the $\lambda_q$ signal using 
3 IMs and 2 phase modulators (PMs) in cascade, see Fig.~\ref{fig:encoding_block}. Within each block of 200~ns, the first 100~ns is carved into a train of 300~ps pulses at 1~ns intervals as the `quantum signal', and each is encoded in both intensity and phase according to the requirements by the TF-QKD protocol. Four intensity levels are used, i.e., $\mu_Z$ (signal state), $\mu_2$ (strong decoy state), $\mu_1$ (weak decoy state) and $\mu_0$ (vacuum state). The subsequent 100~ns in each block is carved into a single square pulse of 70~ns duration and used as the `weak quantum reference'. A 15~ns gap at each end serves as buffer to prevent contamination to the quantum signal. IM1 is used for pulse carving and extinguishing light transmission at vacuum time slots. IM2 produces 4-level intensity modulation and its lowest intensity setting further extinguishes the vacuum signals. IM3 is to set the intensity contrast between the `weak quantum reference' and the quantum signal, with a maximum contrast of up to 700 between their integrated intensities. After intensity modulation, PM1 and PM2 encode the quantum signal pulses with 16 phase slices, $\theta \in \{0, \pi/8, 2\pi/8... 15\pi/8 \}$, to meet the requirement of phase randomization and qubit encoding in TF-QKD protocol. All modulation signals 
are generated by an arbitrary waveform generator (Tektronix, AWG70002B). Our setup has 50~\% transmission duty cycle for the quantum signals, so  the effective QKD clock rate is 500 MHz. We use a 20,000-bits pseudo-random sequence for the quantum signals in the experiment, corresponding to 40~$\mu s$ duration. The encoder is capable of supporting all TF-QKD protocols. 

\subsection{Measurement units}

Charlie has two measurement units (MUs), see Fig.~\ref{active feedback} as well as in Fig.~1(b), Main Text. Coherent MU contains a 50/50 beam splitter, 2 DWDM filters and 3 superconducting nanowire single photon detectors.  The incoming signals from Alice and Bob meet and interfere at the 50/50 beamsplitter, and their interference outcomes are spectrally filtered before detection.
Detectors $D_0$ and $D_1$ are used to derive the raw key from quantum signal $\lambda_q$, with Detector $D_0$ providing also the error signal to drive the fiber stretch (FS) in the coherent dual-band stabilization. Incoherent MU  consists of a polarization beam splitter, a beam splitter, DWDM and inline spectral filters, and 3 single photon detectors. It measures the incoming signals without using interference, and its measurement results are used for delay and polarization drift compensation. 

\subsection{Drifts compensation}

We implement feedback controls to correct for drifts of polarization, phase, optical frequency and temporal arrivals. 

\subsubsection{Active polarization feedback}

Charlie's coherent MU requires the incoming optical signals to have an identical polarization so as to ensure high-visibility interference. To meet this requirement, a polarization beam splitter (PBS) is placed at each incoming fiber path before the 50/50 beam splitter. 
The reflected signal of $\lambda_c$ and $\lambda_q$ by the PBS are routed to detectors $D_2$ and $D_3$, whose count rates are minimised to stabilize the polarization using electrically driven polarization controllers (EPC) placed at Bob (Alice). The error signal is transmitted to Alice and Bob via Internet. The communication latency (several ms) is negligible as compared to the low compensation interval (about 0.5~s). The $\lambda_t$ signal also passes through the reflected path of the PBS, thanks to its polarization set perpendicular in respect to other wavelengths using a manual polarization controller (PC) in the signal path at each transmitter' site (see Fig.~\ref{active feedback}).

\subsubsection{Active phase stabilization}

Coherent dual-band phase stabilization~\cite{zhou2023quantum} we implement has two compensation stages. 
As shown in Fig.~\ref{active feedback}, the fast one compensates for rapid phase drift, typically at several kHz, using the count rate of $D_c$ detector in Charlie's coherent MU as the error signal to drive the PM in Bob's fiber path.  The slow one removes the residual phase drift using detector \textcolor{black}{$D_0$'s} result to adjust the FS in Alice's fiber path.
Two field-programmable-gated-arrays (FPGA1 and FPGA2) handle the two compensation stages with respective feedback rates of 100~kHz and 1~kHz. 

\begin{figure}[ht]
\includegraphics[width=.9\columnwidth]{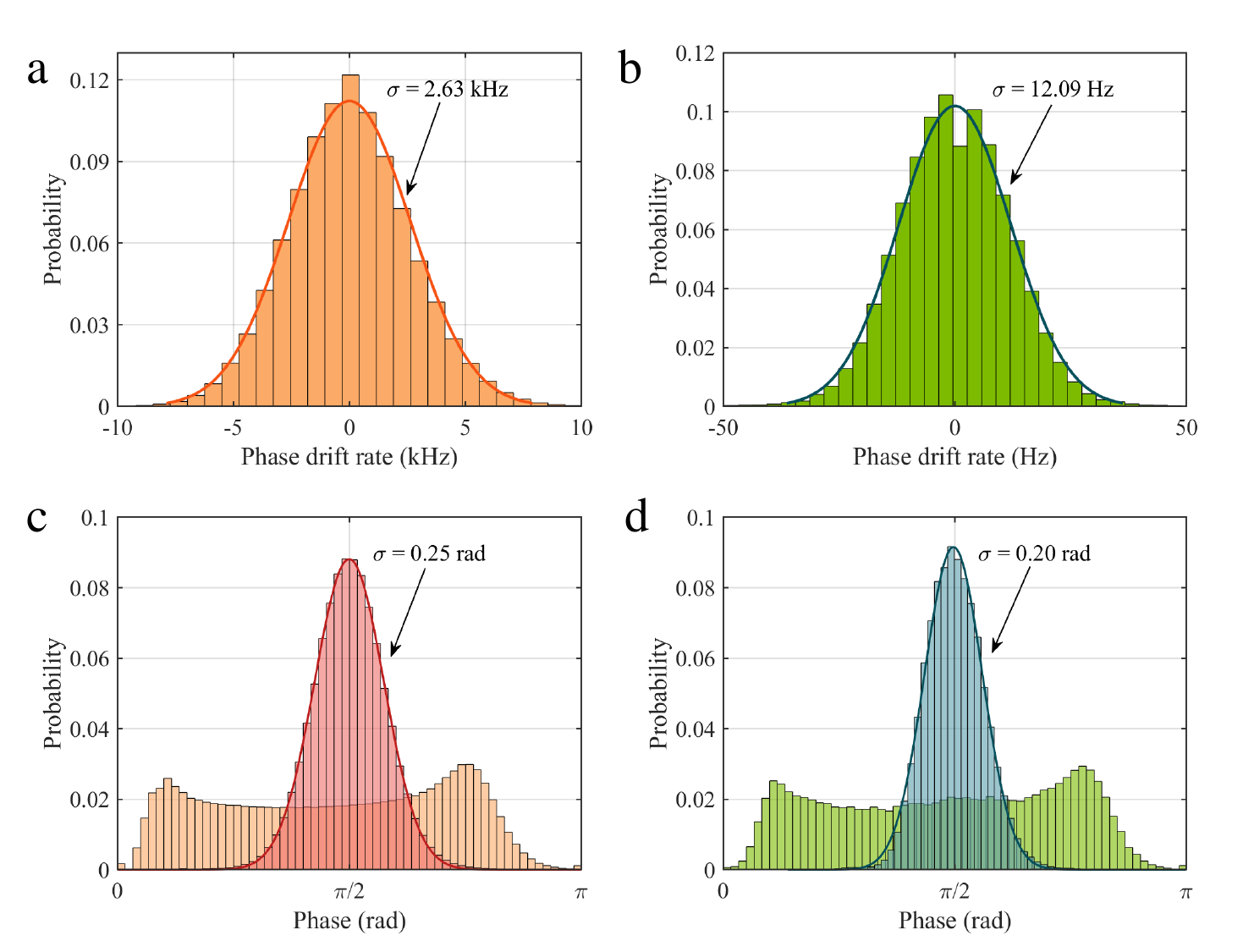}
\caption{Phase drift rate and the stabilization result. \textbf{a}, Histogram of the fast phase drift rate of $\lambda_c$; \textbf{b}, Histogram of the slow residual phase drift rate of $\lambda_q$ after the fast phase compensation; \textbf{c}, The phase angle distribution of the $\lambda_c$ before (orange) and after (red) the channel stabilisation; \textbf{d}, The phase angle distribution of the $\lambda_q$ before (green) and after (blue) the phase stabilisation. All data were measured with a time tagger in 546.61 km quantum fiber.}
\label{phase drift}
\end{figure}

We present a set of stabilisation results obtained over the deployed fiber link in Fig. \ref{phase drift}.  Here, we use Hz (1~Hz = $2\pi$~rad/s) to represent the phase drift rate. Panel~\textbf{a} shows the fast phase drift rate with a standard deviation of 2.63~kHz. Sources for the drift include random phase fluctuation in the hundred of kilometers deployed fiber, optical frequency difference between Alice and Bob's lasers and the finite accuracy of Rubidium clocks used in the OFC units. 
We extract the phase angle of $\lambda_c$ and plot its distribution in Fig.~\ref{phase drift}\textbf{c}, which has a standard deviation of 0.25~rad. With the fast phase compensation on, the phase drift \textcolor{black}{rate} of $\lambda_q$ is substantially slowed down. It has a standard deviation of just 12.09~Hz, as shown in Fig.~\ref{phase drift}\textbf{b}. We then enable FPGA2's proportional-integral-differential (PID) controlling on the fiber stretcher (FS) to compensate the residual phase drift.  Figure~\ref{phase drift}\textbf{d} shows the histogram of angle distribution of the $\lambda_q$ signal to have a standard deviation of 0.20~rad. 

The phase drift rates and the corresponding stabilization results for different fiber links are summarized in Table~\ref{phase drift table}. 
\begin{table}
    \caption{Phase stabilization results for various fiber links.}
    \begin{tabular}{c|p{14mm}|p{15mm}|p{15mm}|p{15mm}}
    \hline
    \hline 
    &Fiber length (km) & Free drift rate (kHz) & Fast phase stabilization (rad) & Slow phase  stabilization (rad) \\
    \hline
    \multirow{2}{*}{Symmetric}
     &546.61 &2.63&0.25 &0.20 \\
      &603.87&2.11 &0.23&0.23 \\
       \hline
      \multirow{1}{*}{Asymmetric}
       &452.46 &2.14&0.24 &0.25 \\
      \hline
      \hline
    \end{tabular}
    \label{phase drift table}
\end{table}

\subsubsection{Frequency compensation}

\begin{figure}[ht]
\includegraphics[width=.9\columnwidth]{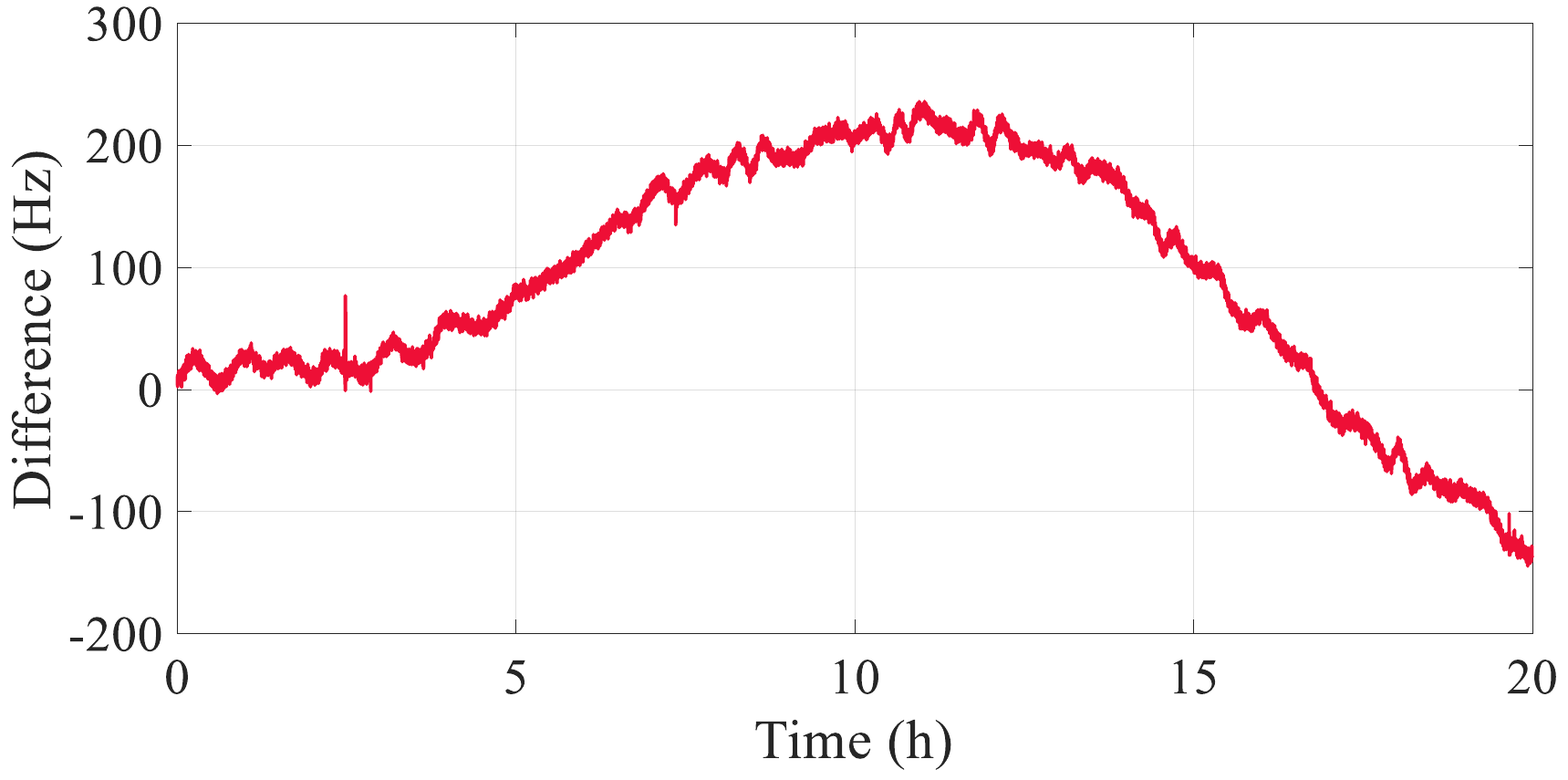}
\caption{Differential frequency between two independent lasers with compensation.}
\label{frequency drift compensated}
\end{figure}

To allow real-time frequency compensation, an acoustic-optic modulator (AOM) is placed between Alice's ultrastable laser and the electro-optic modulator in her OFC unit.
Based on the measurement of the linear frequency drift between the lasers in the users' OFC units  (see Fig.~\ref{frequency drift}), we preset accordingly a compensating shift in the AOM's driving frequency. 
We measure the effect of this frequency pre-compensation in laboratory and the result is shown in Fig.~\ref{frequency drift compensated}.  Within 20~h, the \textcolor{black}{two} lasers' frequency difference stays within $\pm 300$~Hz, a difference that is one order of magnitude smaller than fluctuation caused by the deployed fiber and can thus be corrected for by the coherent dual-band stabilization. 

\subsubsection{Time compensation}

\begin{figure*}
\includegraphics[width=1.9\columnwidth]{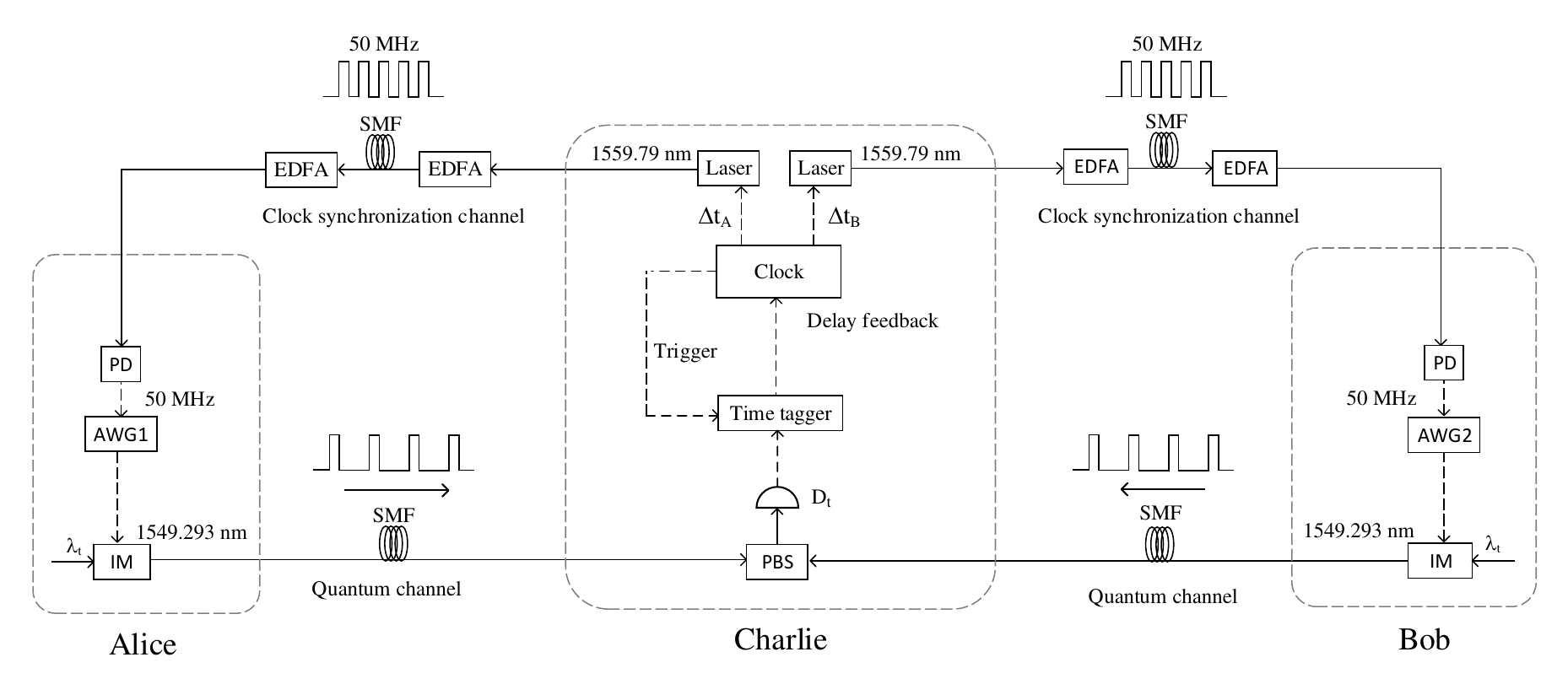}
\caption{Clock synchronization and delay feedback setup. In Charlie the clock is used to synchronize his own time tagger locally. Charlie distributes two 50 MHz optical pulses through clock synchronization channel to synchronize the AWGs in Alice and Bob. AWG: arbitrary waveform generator; SMF: single mode fiber; PBS: polarization beam combiner; PD: photodiode; EDFA: Erbium-doped optical fiber amplifier; SNSPD: superconducting nanowire single photon detector; IM: intensity modulator.} 
\label{fig:clock}
\end{figure*}

To synchronize Alice and Bob's modulation units, Charlie distributes a 50~MHz clock signal to Alice and Bob through the clock synchronization channel, see Fig.~\ref{fig:clock}. 
A total of four erbium doped fiber amplifiers (EDFAs) are used to ensure sufficient optical power reaching the destinations. 
The detailed locations of these EDFAs are as the following, Bogeyan town ($117^\circ62'~$E, $36^\circ39'~$N), Yiyuan City ($118^\circ12'~$E, $36^\circ11'~$N), Zhucheng city ($119^\circ24'~$E, $36^\circ2'~$N) and Huangshan city ($119^\circ99'~$E, $36^\circ17'~$N). 
At each user's site, the 50~MHz optical signal is converted to an electrical clock by a photodiode. 
The delays ($\Delta t_A$ and $\Delta t_B$) of these clocks are adjusted in real time by Charlie based on his measurement of the arrivals of $\lambda_t$ pulses sent by Alice and Bob using the $D_t$ detector in his incoherent MU. This is to ensure optimal alignment between Alice and Bob's quantum pulses and thus high visibility interference.

\subsection{Quantum channel and system loss characterization}

%\clearpage
\begin{table*}
%\internallinenumbers
\caption{Lengths and corresponding losses for the fiber links we used in the experiments.}
\centering
\begin{tabular}{c|c c|c c|c c}
%\toprule
\hline
\hline
\multirow{2}{*}{}& \multicolumn{2}{c}{Total} & \multicolumn{2}{|c}{Alice} &  \multicolumn{2}{|c}{Bob}\\
\cline{2-7}
%\hline
 &length (km) & loss (dB) & length (km) & loss (dB) & length (km) & loss (dB)\\
\hline
\hline
%\midrule
\multirow{2}{*}{Symmetric}
& 546.61 & 100.13 & 273.48 & 50.50 &  273.13 & 49.63\\
& 603.87 & 108.59 & 298.71 & 54.74 & 305.16 & 53.85\\
\hline
\multirow{1}{*}{Asymmetric}
& 452.46 & 84.62 & 248.24 & 46.85 & 204.22 & 37.77 \\
\hline
\hline
%\midrule
%\bottomrule
\end{tabular}
\label{fiberloss}
\end{table*}

The deployed fiber link is formed by ultra-low-loss fiber (G654.E ULL). The buried fiber from Alice (Bob) to Charlie is 223.01~km (204.22~km) with a loss of 42.40~dB (37.77~dB), while the physical separation between Alice and Bob is about 300 km. In Charlie's side, fiber spools (G654.C ultra-low-loss fiber) with a typical loss coefficient of 0.160~dB~km$^{-1}$ are included to vary the length of quantum channel, which is characterised to have an average loss coefficient ranging from 0.180~dB~km$^{-1}$ to 0.187~dB~km$^{-1}$ in the experiments. We summarize the lengths and corresponding losses for the quantum channel in Table~\ref{fiberloss}.

Table~\ref{tab:Charlie_loss} summarizes losses of a series of components with respect to each user's input port in Charlie's module. 
Charlie's transmission loss for Bob's quantum signal is 4.73~dB, which is 0.92~dB higher than Alice's because of the loss by the phase modulator. This loss asymmetry is compensated for with fibers in  the experiments of symmetric configuration. The characteristics of superconducting nanowire single photon detectors (SNSPDs) are presented in Table~\ref{tab:detectors}. 
The detector dark count rate was measured in field under the condition that the $\lambda_c$ ($\lambda_t$) light produces 6~MHz (200 kHz) count rate at detector $D_c$ ($D_t$). The scattered noise from $\lambda_c$ is capped below 1~Hz thanks to the excellent  channel isolation~\cite{zhou2023quantum}.  
For the 603 km fiber experiment, we decrease the bias current in the detectors to further suppress the dark counts at expense of detection efficiencies.

%\clearpage
\begin{table}[ht]
%\internallinenumbers
\caption{Charlie's components loss.}
\centering
%\begin{tabular}{c|p{3cm}|p{3cm}}
\begin{tabular}{c|c|c}
%\toprule
\hline
\hline
 & Alice & Bob\\
\hline\hline
Polarisation beam splitter  & ~~~~~~~0.47~~~~~~~ & ~~~~~~~0.49~~~~~~~\\
\hline
Phase modulator & n/a  &  1.97\\
\hline
Fiber stretch and filter  & 1.05 & n/a \\
\hline
50/50 beam splitter  & 0.66 & 0.65 \\
\hline
DWDM filter & 1.20 & 1.20\\
\hline
Polarisation controller & 0.15 & 0.15\\
\hline
Optical switch & 0.28 & 0.27 \\
\hline
Total loss (dB) & 3.81 & 4.73\\
\hline\hline
%\midrule
%\bottomrule
\end{tabular}
\label{tab:Charlie_loss}
\end{table}

\begin{table*}[t]
%\internallinenumbers
\caption{Characteristics of Charlie's detectors D$_0$ and D$_1$ in different fiber lengths.}
\centering
\begin{tabular}{c|c|c|c|c}
\hline
\hline
  ~~Detector~~ & Efficiency & Dark Count & Efficiency (603 km) & Dark Count (603 km)\\
 \hline\hline
D$_0$ & ~~~~~~~~83~\% ~~~~~~~~& ~~~~~~~~7.80 Hz~~~~~~~~ & 70~\% & 4.15 Hz\\
\hline
D$_1$ & 49~\%  & 1.77 Hz & 47~\% & 1.18 Hz\\
%\hline
%D$_C$  & 80~\% & 10 Hz \\
\hline
\hline
%\bottomrule
\end{tabular}
\label{tab:detectors}
\end{table*}

%\newpage
\section{Detailed experimental parameters and  results}

Table~\ref{tab:parameters} lists the encoding parameters used in the symmetric and asymmetrical experiments. 
Alice and Bob share the same parameters in the symmetric experiment. $P_Z$ ($P_X$) is the probability of encoding a pulse in the Z (X) basis. In the time window of Z basis, weak coherent pulses with signal ($\mu_Z$) and vacuum ($\mu_0$) intensities are randomly prepared with probability $\epsilon$ and $1-\epsilon$. While in the X basis, weak coherent pulses with strong decoy ($\mu_2$), weak decoy ($\mu_1$) and vacuum ($\mu_0$) intensities are randomly prepared with the corresponding probabilities $p_{\mu_2}$, $p_{\mu_1}$  and $p_{\mu_0}$, respectively.  $P_Z + P_X = 1$ and $p_{\mu_2} + p_{\mu_1} + p_{\mu_0} = 1$.

Table~\ref{tab:key rate} summarizes \textcolor{black}{the} experimental results and parameters that are useful for secure key rate (SKR) calculation. $N$ is the total number of transmitted quantum  pulses. The numbers of successful clicks recorded in Charlie are list as ``Detected $\text{AB}_{\text{ab}}$'', where ``A'' (``B'') indicates the X (Z) basis that Alice or Bob uses, and ``a'' (``b'') is 0, 1, 2 or 3 representing the chosen intensity $\mu_0$, $\mu_1$, $\mu_2$ or $\mu_Z$. QBER ($\text{X}_{11}$) and QBER ($\text{X}_{22}$) indicate the error rate when Alice and Bob both send decoy state $\mu_1$ and $\mu_2$ in the X basis, respectively. SKC$_0$ is the absolute repeaterless Pirandola-Laurenza-Ottaviani-Banchi (PLOB) bound. The rest of the notations see the Section I.

\begin{table}[ht]
%\internallinenumbers
\caption{Encoding parameters used in the symmetric and asymmetrical setups.}
\centering
\begin{tabular}{c|c|c|c|c}
\hline\hline
\multirow{3}{*}{Parameter}{} & \multicolumn{2}{c|}{Symmetric} & \multicolumn{2}{c}{  Asymmetric }  \\
\cline{2-5}
 &  \quad546.61km \quad& \quad603.87km \quad& \multicolumn{2}{c}{ \quad \quad452.46km \quad \quad}\\
\cline{2-5}
 & Alice/Bob & Alice/Bob &  Alice  &  Bob  \\
\hline
$\mu_Z$  & 0.493 & 0.423 &  0.493& 0.247 \\
$\mu_2$  & 0.493 & 0.252 &  0.493& 0.077 \\
$\mu_1$  & 0.090 & 0.056 &  0.113& 0.018 \\
$\mu_0$  & 0.0002 & 0.0002 & 0.0002 & 0.0002\\
\hline
$P_Z$ & \multicolumn{2}{c|}{0.735} & ~~0.735~~ & 0.735 \\
$P_X$ %= 1 - P_Z$
& \multicolumn{2}{c|}{0.265} & 0.265 & 0.265\\
$\epsilon$ & \multicolumn{2}{c|}{0.269} & 0.405 & 0.141\\
$p_{\mu_2}$ & \multicolumn{2}{c|}{0.316} & 0.316 & 0.316\\
$p_{\mu_1}$ & \multicolumn{2}{c|}{0.606} & 0.606 & 0.606\\
$p_{\mu_0}$ %= $1 - p_{\mu_2} -p_{\mu_1} $
& \multicolumn{2}{c|}{0.078} & 0.078 & 0.078 \\
\hline\hline
\end{tabular}
\label{tab:parameters}
\end{table}
 
\begin{table*}[t]
\caption{Experimental results at various quantum link fiber lengths with SNS-AOPP-TF-QKD protocol.}
\centering
\setlength{\tabcolsep}{3mm}{
\begin{tabular}{c | c c | c }
\hline \hline
Total length (km)  & 546.61 & 603.87  & 452.46\\
\hline
Alice-Charlie (km)  & 273.48 & 298.71  & 248.24\\
Bob-Charlie (km)  & 273.13 & 305.16  & 204.22\\
 \hline
$N$  & $2.772 \times 10^{13}$ & $4.05 \times 10^{12}$ & $4.28 \times 10^{12}$ \\
% Number of phase slices & 16 & 16 &16 \\
\hline
Detected $\text{XX}_{20}$  & 68859 & 1856 & 23105\\
Detected $\text{XX}_{02}$  & 67693 & 2162 & 21269\\
Detected $\text{XX}_{10}$  & 29156 & 878 & 9097\\
Detected $\text{XX}_{01}$  & 24569 &  1163 & 8878 \\
Detected $\text{XX}_{00}$  & 63 &   7 &  21\\
Detected $\text{XZ}_{00}$  & 1847 &  195 & 363 \\
Detected $\text{XZ}_{10}$  & 649804 &  24830 & 198482 \\
Detected $\text{XZ}_{20}$  & 1805364 &  60929 & 443226 \\
Detected $\text{ZX}_{00}$  & 1924 &  220 &  496\\
Detected $\text{ZX}_{01}$  & 621437 &  26334 &  320369\\
Detected $\text{ZX}_{02}$  & 1787060 &   57269 & 707987\\
Detected $\text{ZZ}_{03}$  & 4005761 &  234768 &  2556562\\
Detected $\text{ZZ}_{30}$  & 4396652 &  245490 & 1708076 \\
Detected $\text{ZZ}_{33}$  & 3107361 &  177447 & 1549556 \\
Detected $\text{ZZ}_{00}$ & 51305 &  5757 & 10474 \\
\hline
QBER ($\text{X}_{11})$  & 8.71\% & 8.50\% & 6.87\%\\
QBER ($\text{X}_{22})$   & 9.63\% & 8.82\% & 6.75\%\\
\hline
QBER ($E_z$ before AOPP)  &  27.32\% & 27.61\% & 26.38\%\\
QBER ($E_z$ after AOPP)  & 0.90\% & 1.74\%  & 0.37\%\\
$n_1$ (Before AOPP)  & $4.901 \times 10^6$ & $2.932 \times 10^5$ & $2.774 \times 10^6$\\
$n_1$ (After AOPP) & $7.995 \times 10^5$ & $5.061 \times 10^4$ & $5.893 \times 10^5$ \\
$e_1^{ph}$ (Before AOPP) & 11.28\% & 7.90\% & 9.94\%\\
$e_1^{ph}$ (After AOPP)  & 20.01\% & 14.55\% & 17.90\%\\
\hline
SKR (bit/s) - finite size  & 0.53 & / & 16.06 \\
SKR (bit/signal) - finite size  & $1.060 \times 10^{-9}$& / & $3.212 \times 10^{-8}$\\
SKR (bit/s) - asymptotic  & / & 0.12 & / \\
SKR (bit/signal) - asymptotic  & / & $2.455 \times 10^{-10}$ & / \\

SKC$_0$ (bit/signal)  &  $1.400 \times 10^{-10}$ & $1.996 \times 10^{-11}$ & $4.979 \times 10^{-9}$\\
Ratio SKR over SKC$_0$  & 7.57 & 12.30 & 6.45 \\

\hline \hline

\end{tabular}
}
\label{tab:key rate}
\end{table*}

%\bibliography{TFQKDbib}


\end{document}